\documentstyle[twoside,fleqn,espcrc2,epsfig]{article}

\title{Progress in understanding colour confinement.}

\author{A. Di Giacomo
\address{Dipartimento di Fisica Universit\`a and INFN
Pisa (Italy)}
\thanks{Presented the talk.}
, B. Lucini$^a$, L. Montesi$^a$, G. Paffuti$^a$
}

\begin{document}

\begin{abstract}
New results from lattice are presented, which demonstrate that monopoles
condense in the vacuum of the confined phase of QCD, which is thus a dual
superconductor.

Monopoles defined by different abelian projections appear to be physically
equivalent.
\end{abstract}

\maketitle

\section{Introduction}
An appealing mechanism for confinement of colour in QCD is dual
superconductivity of the vacuum\cite{1,2}. By dual Meissner effect the
chromoelectric field acting between a $q\,\bar q$ pair is confined into an
Abrikosov flux tube, which in the ground state configuration has energy
proportional to the distance: $V(r) = \sigma r$. $\sigma$ is the string tension.

The idea that the strings appearing in hadron phenomenology could be Abrikosov
flux tubes goes back to ref.\cite{3}, which was inspired by the Veneziano
model\cite{4}.

Due to asymptotic freedom, it is very likely that QCD exists as a field theory,
or that its euclidean version describes a statistical system for which a
thermodynamical limit exists. For the same reason a significant sample of
lattice configurations should be a good approximation to that limit and should
identify the true vacuum, if the correlation length $\lambda$ is much larger
than the lattice spacing $a$, and much smaller then the lattice size $L$
($a \ll \lambda \ll L$).

Dual superconductivity means condensation of monopole charges, or spontaneous
breaking of the $U(1)$ symmetry related to their conservation.

The colour deconfining transition is therefore, in this mechanism, a
transition order-disorder, which can be investigated as a change of symmetry by
use of an order parameter\cite{5}.

A sensible strategy to attack the problem is
\begin{itemize}
\item[a)] identify the relevant magnetic $U(1)$
\item[b)] detect the deconfining transition by looking at the symmetry of the
vacuum.
\end{itemize}
The question a) will be discussed in sect.2, the question b) in sect.3.
\section{Monopoles in non abelian gauge theories.}
The prototype of monopoles are t'Hooft\cite{6}-Polyakov\cite{7} monopoles,
which were discovered as solitons of the Georgi Glashow model\cite{8}
\begin{equation}
{\cal L} = -\frac{1}{4} \vec G_{\mu\nu}\vec G_{\mu\nu} + D_\mu\vec\Phi
D_\mu\vec\Phi - V(|\vec \Phi|) \label{eq:1a}\end{equation}
\begin{equation}
V(|\vec \Phi|) = \frac{\lambda}{4}\left(\vec \Phi^2 -
\frac{m^2}{\lambda}\right)^2 \label{eq:2}\end{equation}
If $m^2 > 0$, the $SO(3)$ symmetry of the model spontaneously breaks \`a la
Higgs to $U(1)$, and $\vec\Phi_0\equiv\langle\vec \Phi\rangle\neq 0$.

A time independent ansatz of the form
\[
\vec \Phi(\vec r) = f(r) \vec\Phi_0 \hat r \quad
A_0^a = 0\quad A^a_i = \frac{h(r)}{g}\varepsilon_{iak}\frac{r^k}{r}
\]
with $f(r), h(r)\to1$ as $|\vec r|\to \infty$, admits a solution with finite
energy, in which $f(r)$, $g(r)$ are practically equal to 1 everywhere except in
a small radius $a\sim 1/\mu$. The solution has the geometry of a monopole.
This is explicitly seen by transforming to the gauge in which $\hat\Phi \equiv
\vec\Phi/|\vec \Phi|$ is constant, say $\hat\Phi = (0,0,1)$: the
corresponding gauge transformation $U(\vec r)$ is called abelian projection.
$U(\vec r)$ is singular at $\vec r=0$, where $\vec \Phi=0$.

The abelian gauge field of the residual $U(1)$ symmetry in this
gauge
\begin{equation}
{\cal F}_{\mu\nu} = \partial_\mu A_\nu^3 - \partial_\nu
A^3_\mu\label{eq:3}\end{equation}
is the field of a Dirac monopole in the soliton configuration
\begin{equation}
\vec E = 0\qquad \vec H \simeq_{r\to\infty} \frac{2}{g}\frac{\hat r}{r^2} +
\hbox{Dirac string}\label{eq:4}\end{equation}
In a covariant form
\begin{equation}
{\cal F}_{\mu\nu} = \hat\Phi\vec G_{\mu\nu} - \frac{1}{g}
\hat\Phi(D_\mu\hat\Phi\wedge D_\nu\hat\Phi)\label{eq:5}\end{equation}
In terms of
${\cal F}^*_{\mu\nu} = \frac{1}{2}\varepsilon_{\mu\nu\rho\sigma} {\cal
F}_{\rho\sigma}$
the magnetic current is defined by $\partial_\mu{\cal F}^*_{\mu\nu} = j^M_\nu$
and is identically conserved:
\begin{equation}
\partial_\mu j^M_\mu = 0\label{eqn}
\end{equation}
A magnetic $U(1)$ symmetry exists. Both ${\cal F}_{\mu\nu}$ and the magnetic
charge $Q$ are colour singlets.

One can operate the abelian projection
\begin{equation}
U(\vec r) \hat\Phi(\vec r) = (0,0,1)\label{eq:6}\end{equation}
on generic configurations also in the unbroken phase of the system, where
monopoles do not exist as solitons. $U(\vec r)$ will be singular at the sites
where
$|\vec\Phi| = 0$ and the singularities will reflect in the topology of the
gauge field, as a Dirac strings.

Monopoles are lumps in the Higgs phase, but they can also exist in the
symmetric phase, where they condense.

In QCD there is no Higgs field. However any operator in the adjoint
representation can play the role of $\vec\Phi$, and define monopoles which are
exposed by the corresponding abelian projection.

On the lattice any closed path of parallel transport defines a $\vec\Phi$ in
the adjoint representation. An open plaquette, an open Polyakov line, an open
``butterfly'',
$\varepsilon_{\mu\nu\rho\sigma}\Pi_{\mu\nu}(n)\Pi_{\rho\sigma}(n)$. There exist
in fact a continuous infinity\cite{9}
of monopoles species. What monopoles do condense in the
vacuum? A guess of t'~Hooft\cite{9} is that all of them are physically
equivalent: configuration by configuration the location and the number
of the monopoles is different, but on the average they are indistinguishable.

There is a community of practitioners of the maximal abelian
projection\cite{10} or of the Delambertian projection\cite{11}, who believe that their choice
is better than others because
of abelian and monopole dominance\cite{12}. The physical quantities, like the
string tension, relevant to confinement, are indeed approximated within $20\%$
by the corresponding quantities computed in the abelian projected $U(1)$, or in
terms of the monopole part of them,  meaning that the projection
identifies the degrees of freedom relevant to confinement.

\section{The disorder parameter for dual superconductivity.}
The basic idea is to construct an operator $\mu$ which carries non zero
monopole charge, and use its vacuum expectation value $\langle\mu\rangle$ as a
detector of symmetry. $\langle\mu\rangle\neq 0$ signals spontaneous breaking
of magnetic $U(1)$, and hence, under very general assumptions, dual
superconductivity. In a theory in which electric charges are pointlike,
magnetic monopoles have non trivial topology, due to Dirac string: moreover
they are coupled with magnetic charge $m = n/e$ due to Dirac quantization
condition.

A similar feature is present in  a variety of systems in statistical mechanics
and is known as duality. The prototype is the 2d Ising model\cite{13},
where the
system can be described either in terms of local spin $\sigma_i = \pm1$, or in
terms of 1 dimensional kinks which are spins on the dual
lattice. The partition function is the same with the
correspondence $\beta\to\sim1/\beta$ which maps strong coupling regime of the
model with weak coupling of its dual.

The basic idea to construct a creation operator for a topological
configuration is to shift the field configuration by the classical topological
configuration by the translation operator\cite{13,14,15,5}.

In the same way as
$ e^{ip a} |x\rangle = |x + a\rangle $

\begin{equation}
\mu(\vec y,t) = \exp\left[i\int\Pi(\vec x,t)\Phi_{cl}(\vec x,\vec y) d^3
x\right]\label{eq:7}\end{equation}
operates on a field configuration $|\Phi(\vec x,t)\rangle$ as
\[\mu(\vec y,t)|\Phi(\vec x,t)\rangle =
|\Phi(\vec x,t) + \Phi_{cl}(\vec x,t)\rangle\]
Technical modifications are needed with compact theories, where the field cannot
be shifted arbitrarily\cite{5a}.

What is measured on the lattice is the correlator
\begin{equation}
{\cal D}(t) = \langle\bar\mu(\vec 0,t) \mu(\vec 0,0)\rangle
\label{eq:8}\end{equation}
which describes a monopole sitting at site $\vec x = 0$ and propagating from
time $x_0=0$ to $x_0=t$.

At large values of $t$
\begin{equation}
{\cal D}(t) \simeq A\exp(-M t) + \langle\mu\rangle^2 \label{eq:9}\end{equation}
$\langle\mu\rangle\neq0$ means dual superconductivity.

At finite temperature a direct measurement of $\langle\mu\rangle$ can be done:
then antiperiodic boundary conditions in time are needed\cite{5a}.

Instead of ${\cal D}(t)$ it can prove numerically convenient to measure
$\rho(t) = \frac{1}{2}\frac{d}{d\beta}\ln{\cal D}(t)$.

At large values of $t$ again from eq.(\ref{eq:9})
\begin{equation}
\rho(t)\simeq C \exp(-M t) + \rho\label{eq:10}\end{equation}
with $\rho = \frac{d}{d\beta}\ln\langle\mu\rangle$. Since $\langle\mu\rangle =
1$ for $\beta = 0$
\begin{equation}
\langle\mu\rangle = \exp(\int_0^\beta\rho(\beta') d \beta')
\label{eq:11}\end{equation}
The typical
behaviour  of $\rho$ at the deconfining phase
transition for $SU(2)$ gauge theory is shown in fig.1, for different
monopole species.
\par\noindent
\begin{minipage}{0.95\linewidth}
\epsfxsize0.95\linewidth
{
\epsfbox{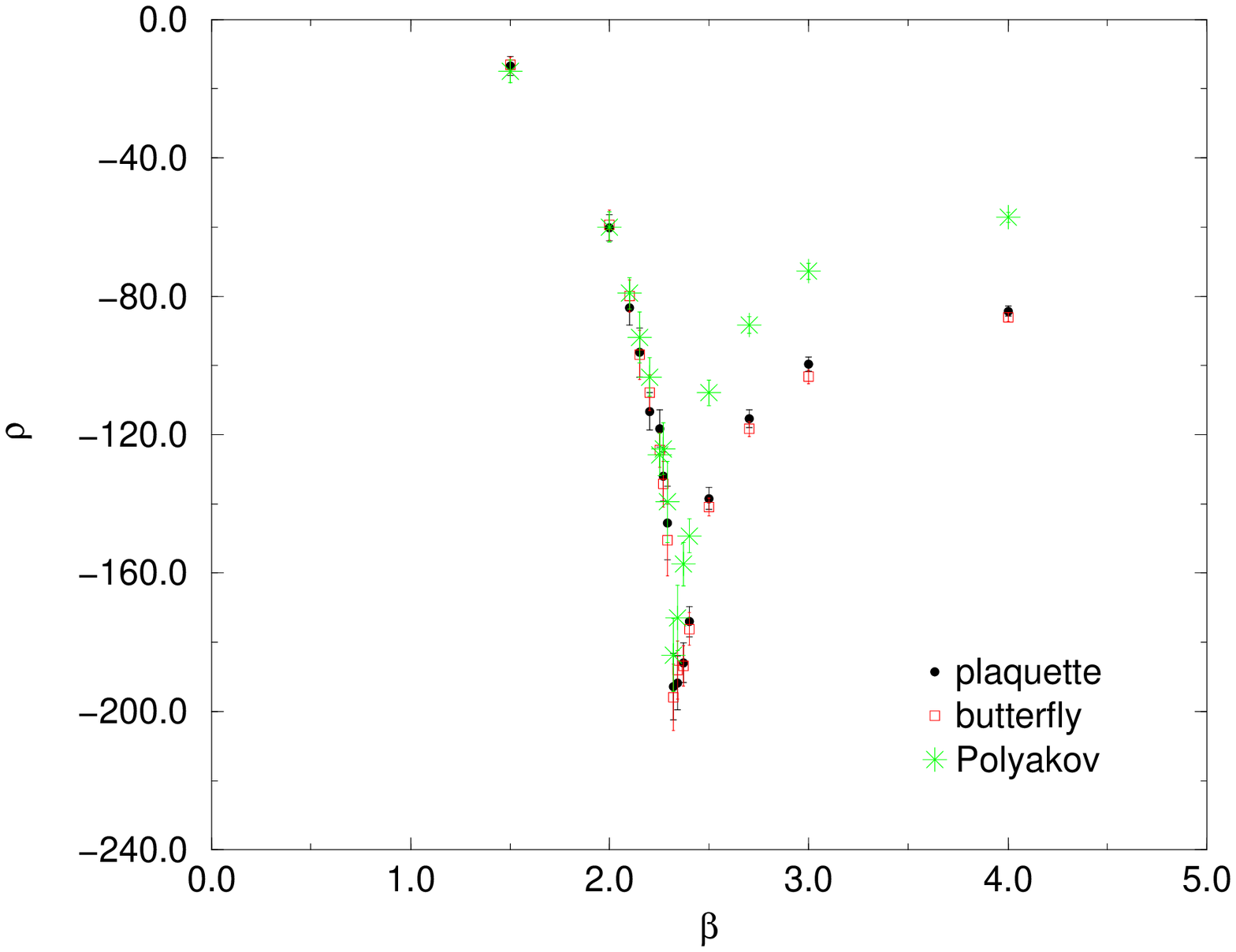}}
\end{minipage}\\
{\bf Fig.1} $\rho$ vs $\beta$ for different abelian projections
in $SU(2)$. The negative peak signals phase transition.
\vskip0.1in\par\noindent
The main results are
\begin{itemize}
\item[a)] Different monopole species have indistinguishable behaviour.
\item[b)] For $\beta < \beta_c$ $\rho$ tends to a finite limit as spatial
volume goes to $\infty$, showing that $\langle\mu\rangle\neq 0$ in the
confined phase.
\item[c)] For $\beta > \beta_c$ the regime is perturbative and
\[ \rho\sim - |c| L_S + c' \]
with $L_S$ the space extension of the lattice. This means that
$\langle\mu\rangle\to 0$ as $L_S\to \infty$, i.e. in the thermodynamical limit.
The expectation for $\langle\mu\rangle$ as a disorder parameter would be zero
for $\beta> \beta_c$. This can only happen in the thermodynamical limit: for
finite volume $\langle\mu\rangle$ is  an analytic function of $\beta$ for any
finite size of the lattice, and cannot vanish
identically for $\beta > \beta_c$, except if it vanishes
for all values of $\beta$. Only in the infinite volume limit Lee Yang
singularities develop\cite{16} and $\langle\mu\rangle$ can become a real
disorder parameter\cite{5a}.
\item[d)] For $\beta\sim\beta_c$ $\rho$ has a sharp negative peak, which means
an  abrupt decrease of $\langle\mu\rangle$ towards $0$. In this region the
correlation length goes large,
\begin{equation}
\xi\simeq (\beta_c-\beta)^{-\nu}\label{eq:12}\end{equation}
with $\nu$ a critical index. By dimensional arguments,
\begin{equation}
\langle\mu\rangle =
f(\frac{a}{\xi},\frac{L}{\xi})\mathop\simeq_{\beta\to\beta_c}
f(0,\frac{L}{\xi})\label{eq:13}\end{equation} or by eq.(\ref{eq:12})
\begin{equation}
\langle\mu\rangle = F(L^{1/\nu}(\beta_c-\beta))\label{eq:14}\end{equation}
whence the scaling law follows
\begin{equation}
\rho/L^{1/\nu} = \Phi(L^{1/\nu}(\beta_c-\beta))\label{eq:15}
\end{equation}
Scaling is obeyed only with the proper value of index $\nu$, which can be
then determined, together with $\beta_c$. The quality of scaling is shown in
fig.2 and the corresponding value of $\nu$, determined by best fit
procedure is
\begin{equation}
\nu = 0.62 \pm .02\label{eq:16}\end{equation}
The expectation is that the transition is second order and that it belongs to
the same class of universality as the 3d Ising model, or $\nu = .631(1)$
\end{itemize}
\par\noindent
\begin{minipage}{0.95\linewidth}
\epsfxsize0.95\linewidth
{
\epsfbox{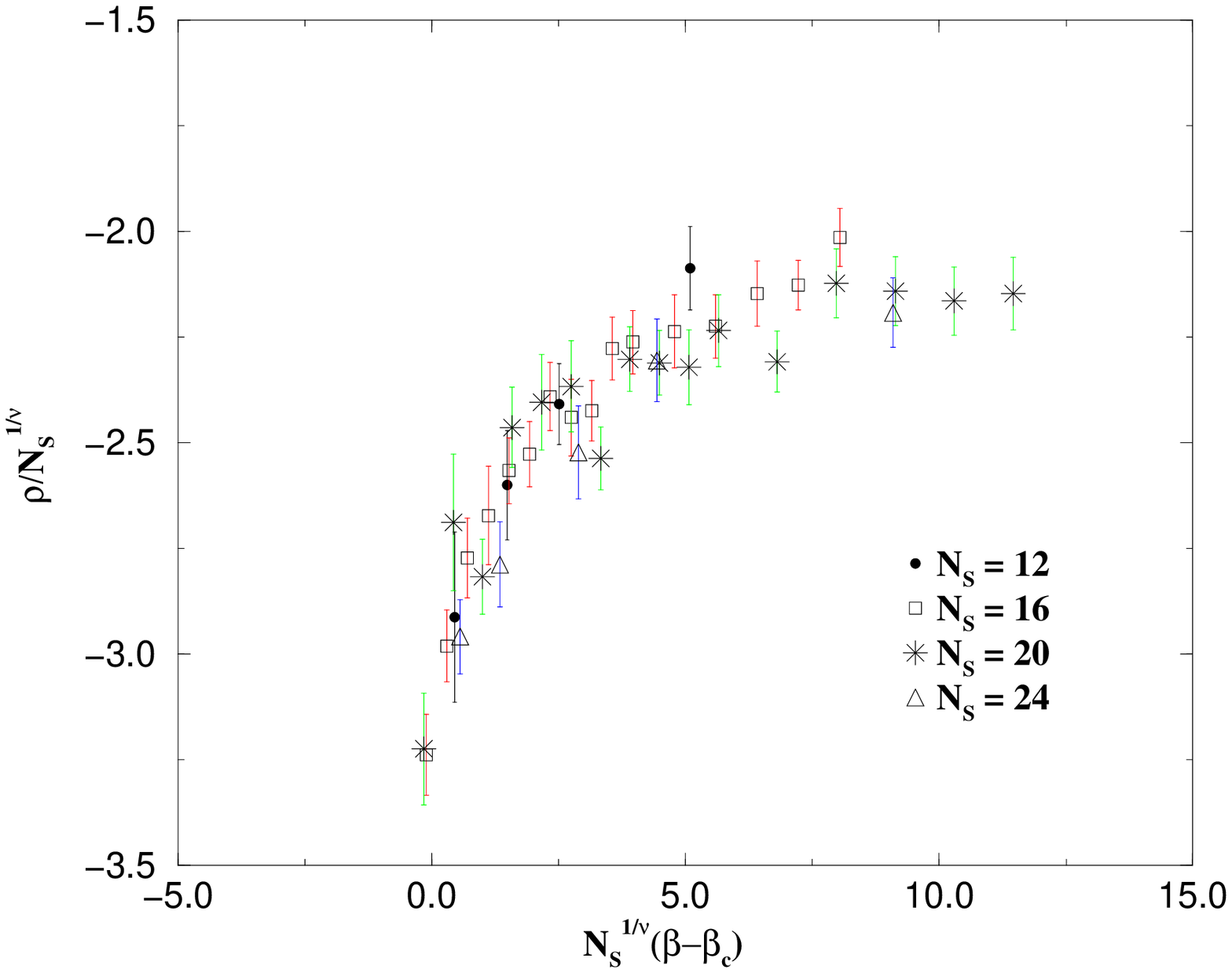}}
\end{minipage}\\
{\bf Fig.2} Finite size scaling eq.(\ref{eq:15}). $SU(2)$.
\vskip0.1in\par\noindent
A similar analysis\cite{17} for $SU(3)$ shows that there is no appreaciable
difference neither
in the disorder parameters for the 2 possible choices of monopoles
defined by a given abelian projection, nor  between
different abelian projections (plaquette, Polyakov line, butterfly),
fig.3,fig.4.
\par\noindent
\begin{minipage}{0.95\linewidth}
\epsfxsize0.95\linewidth
{
\epsfbox{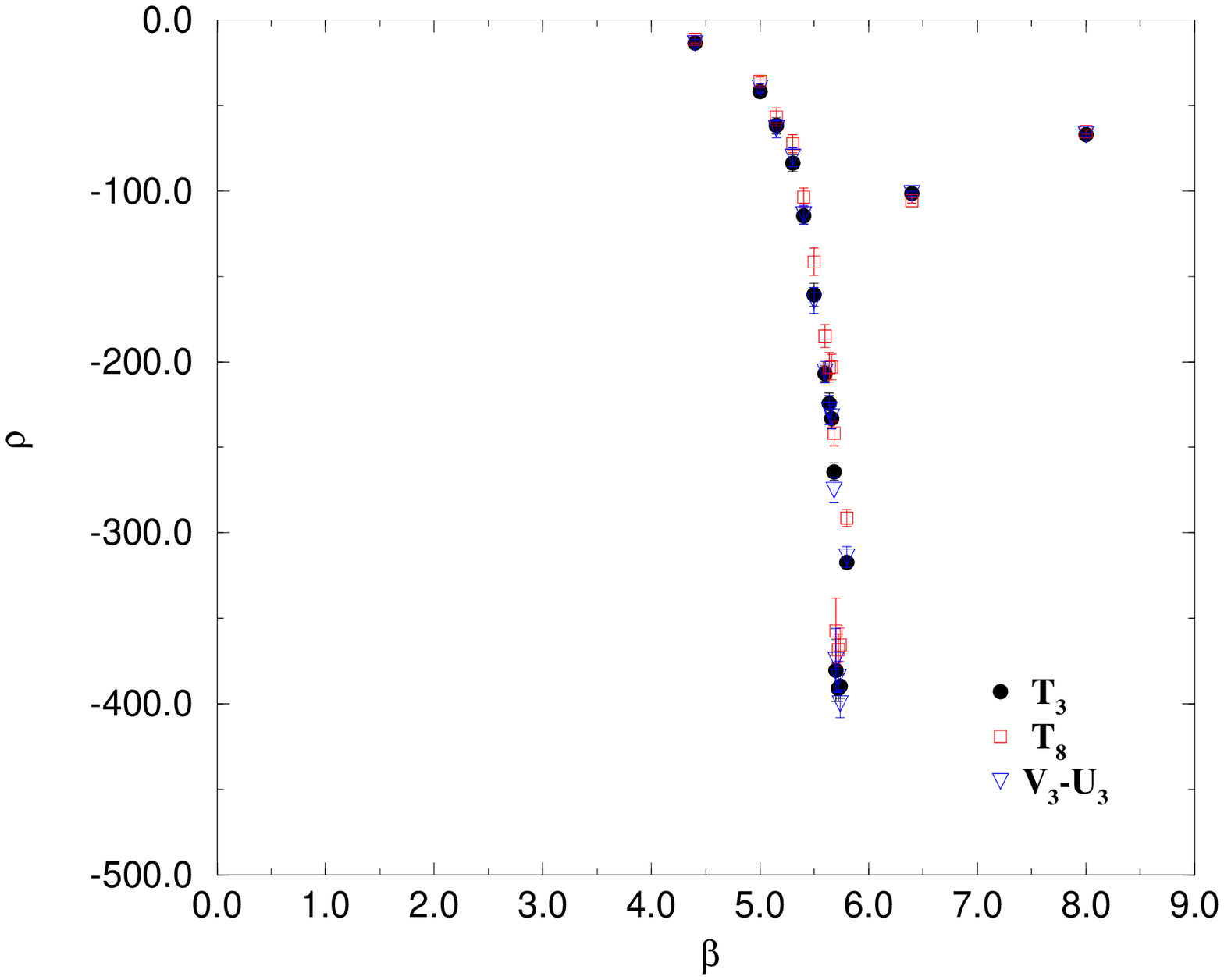}}
\end{minipage}\\
{\bf Fig.3} $\rho$ for the two different monopole species
in the Polyakov projection. $SU(3)$.
\vskip0.1in\par\noindent
An attempt to determine an effective critical index for the transition, which
is known to be weak first order gives $\nu \sim 0.6$.

The expectation for large enough values should be $1/3$, or $1/d$. Larger
volumes are under investigation to check that.
\par\noindent
\begin{minipage}{0.95\linewidth}
\epsfxsize0.95\linewidth
{
\epsfbox{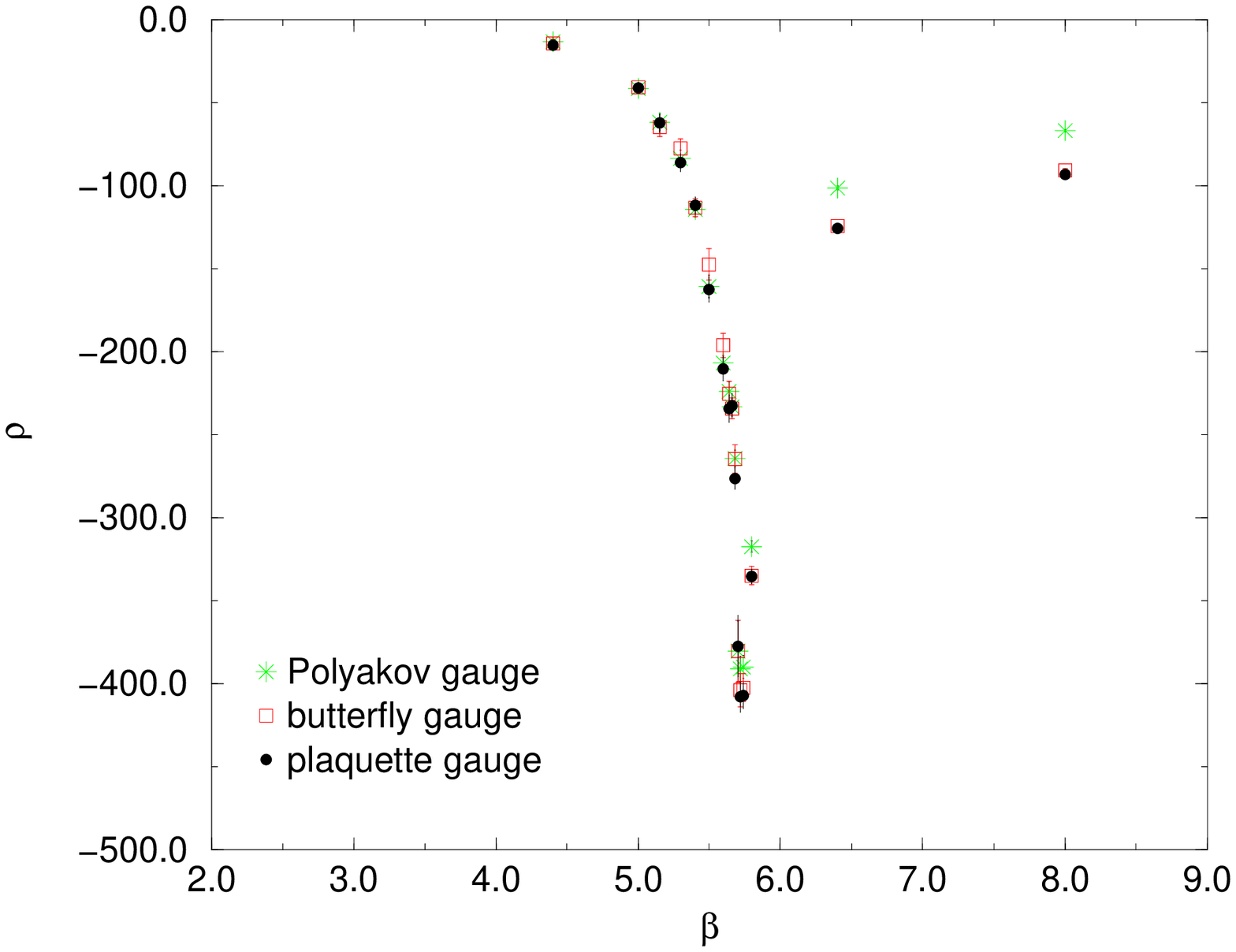}}
\end{minipage}\\
{\bf Fig.4} $\rho$ for different abelian
projections. $SU(3)$.
\vskip0.1in\par\noindent

\section{Conclusions.}
Vacuum of $SU(2)$, $SU(3)$ non abelian gauge theories
in the confined phase
is a dual
superconductor, independent of the abelian projection used\cite{9}. This
evidence comes from direct investigation of the symmetry of the vacuum,
by a disorder parameter detecting monopole condensation. The critical index
for $SU(2)$ deconfining phase transition
can be determined and is
compatible with 3d Ising model, as
expected.
\section*{Acknowledgements}
Partially supported by MURST and by EC TMR program
         ERBFMRX-CT97-0122

\end{document}